\documentclass{ws-ijmpe}

\begin{document}

\markboth{V.K. Magas, L.P. Csernai, E. Molnar}{Bjorken expansion with gradual freeze out}

\catchline{}{}{}{}{}
\renewcommand{\vec}[1]{\mathbf{#1}}
\def\be{\begin{equation}}
\def\ee{\end{equation}}
\def\bea{\begin{eqnarray}}
\def\eea{\end{eqnarray}}
\newcommand{\beq}{\begin{equation}}
\newcommand{\eeq}[1]{\label{#1} \end{equation}}

\title{Bjorken expansion with gradual freeze out}

\author{\footnotesize V.K. Magas}

\address{Departament d'Estructura i Constituents de la Mat\'eria, \\
Universitat de Barcelona, Diagonal 647, 08028 Barcelona, Spain\\
        vladimir@ecm.ub.es}

\author{L.P. Csernai}

\address{Theoretical and Energy Physics Unit, University of Bergen, \\
        Allegaten 55, 5007 Bergen, Norway $\&$\\
        MTA-KFKI, Research Inst of Particle and Nuclear Physics, \\
        H-1525 Budapest 114, P.O.Box 49, Hungary\\
        csernai@ift.uib.no}

\author{E. Molnar}

\address{Frankfurt Institute for Advanced Studies,
Johann Wolfgang Goethe University\\
Max-von-Laue-Str. 1,
60438 Frankfurt am Main,
Germany\\
molnar@fias.uni-frankfurt.de}

\maketitle

\begin{history}
\received{(received date)}
\revised{(revised date)}
\end{history}

\begin{abstract}
The freeze out of the expanding systems, created in relativistic heavy ion collisions, will be discussed. We combine kinetic  freeze out equations with Bjorken type system expansion into a unified model. Such a model is a more physical generalization of the earlier simplified non-expanding freeze out models. We shall see that the basic freeze out features, pointed out in the earlier works, are not smeared out by the expansion. 
\end{abstract}

\section{Model for the simultaneously expanding and freezing out fireball}
In this paper we present a simple freeze out (FO) model, which describes the FO of
particles from a Bjorken expanding fireball \cite{Bjorken}. The important feature of the proposed scenario is that physical freeze out is 
completely finished in a finite time, which can be varied from $0$ (FO hypersurface) to 
$\infty$. In the other words our FO happens in a layer, i.e. in a domain restricted by two parallel 
hypersurfaces $\tau=\tau_1$ and $\tau=\tau_1+L$, where $\tau$ is the proper time variable and $L$ is
the maximal duration of the FO process.
\\ \indent
The present model simultaneously describes the gradual FO  and the expansion of the system, 
and thus it is a more realistic extension of the oversimplified
FO models, which did not include system expansion \cite{old_SL_FO,old_TL_FO,Mo05a,Mo05b}.
In Ref. \cite{old_TL_FO} authors have also adopted kinetic gradual FO model to Bjorken geometry, but combined it with Bjorken expansion on the consequent, not on the parallel basis: system expands according to Bjorken hydro scenario, but when it reaches beginning of the FO process system stops expansion and gradually freezes out in a fixed volume. It was shown in \cite{old_TL_FO} that although such a model allows to obtain analytical results, it is not physical, the simultaneous modeling of expansion and freeze out is required in order to avoid decreasing of the total entropy. And now we propose such a generalized model \cite{Bjorken_FO}.
\\ \indent
We start with introducing two components of the distribution function, $f$: the interacting, $f^i$, and the frozen out,
$f^f$  ones, ($f=f^i+f^f$), then, correspondingly we will have two components of the energy density and baryon density. Details of the derivation can be found in Ref. \cite{Bjorken_FO} and here we only give the final coupled system of equations 
for interacting and free components describing the change in the particle density and energy density:
\beq
\frac{d e^i}{d \tau}=-\frac{e^i+P^i}{\tau}-\frac{e^i}{\tau_{FO}}\frac{L}{L+\tau_1-\tau}\, ;
\quad \frac{d n^i}{d \tau}=-\frac{n^i}{\tau}-\frac{n^i}{\tau_{FO}}\frac{L}{L+\tau_1-\tau}\,,
\eeq{int}
\beq
\frac{d e^f}{d \tau}=-\frac{e^f}{\tau}+\frac{e^i}{\tau_{FO}}\frac{L}{L+\tau_1-\tau}\, ;
\quad \frac{d n^f}{d \tau}=-\frac{n^f}{\tau}+\frac{n^i}{\tau_{FO}}\frac{L}{L+\tau_1-\tau}\,.
\eeq{free}

\indent
All together we have the following simple model describing the evolution of the fireball created in relativistic heavy 
ion collision. 
\\
{\bf Initial state: $\tau=\tau_0$}\, ,
\beq
e(\tau)=e_0\,, \quad
n(\tau)=n_0\,.
\eeq{phase0}
{\bf Phase I, Pure Bjorken hydrodynamics: $\tau_0\le \tau\le \tau_1$}\, ,
\beq
e(\tau)=e_0\left(\frac{\tau_0}{\tau}\right)^{1+c_0^2}\,, \quad
n(\tau)=n_0\left(\frac{\tau_0}{\tau}\right)\,,
\eeq{pure_bjor}
where $P=c_0^2 e$ is the equation of state (EOS).
\\
{\bf Phase II, Bjorken expansion and gradual FO:  $\tau_1\le \tau\le \tau_1+L$} \, ,
\beq
e^i(\tau)=e_0\left(\frac{\tau_0}{\tau}\right)^{1+c_0^2} \left(\frac{L+\tau_1-\tau}{L}\right)^{L/\tau_{FO}},
\,
n^i(\tau)=n_0\left(\frac{\tau_0}{\tau}\right)\left(\frac{L+\tau_1-\tau}{L}\right)^{L/\tau_{FO}}.
\eeq{bjor_FO_2}
The difference with respect to the pure Bjorken solution eqs. (\ref{pure_bjor}) is the multiplier describing the gradual FO 
of the system. 
We also see that the interacting component will vanish when we reach the end of the FO layer, i.e. $\tau \rightarrow L+\tau_1$.

\begin{figure}[htb!]
\centering
\includegraphics[width=0.9\textwidth]{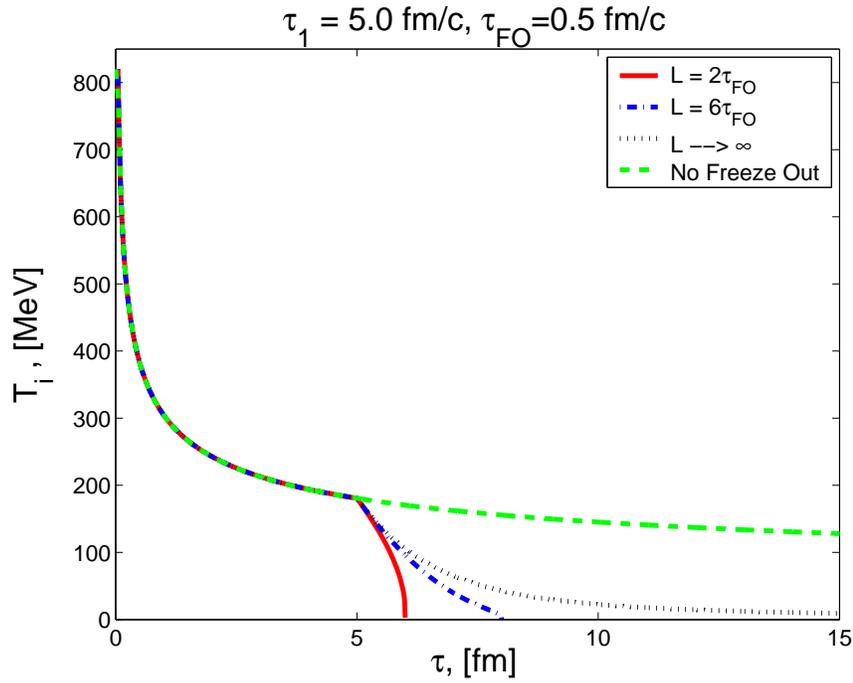}
\caption{Evolution of temperature of the interacting matter for
different FO layers. $T_i(\tau_0=0.05\ fm)=820\ MeV$, $T_{FO}=180\ MeV$. "No Freeze Out" means that we used standard Bjorken hydrodynamics even in phase II.}
\label{fig1}
\end{figure}

\indent
Knowing $e^i(\tau)$ and equation of state (EoS) we can calculate the temperature of the interacting component, $T_i(\tau)$, 
as a function of the proper time.
Due to symmetry of the system, $u_i^\mu(\tau)=u^\mu(\tau_0)=(1,0,0,0)$. 
Thus, we have complete knowledge about the evolution of the interacting component, $f^i(\tau)$, which is a thermal distribution with given $T_i(\tau)$, $n^i(\tau)$, $u_i^\mu(\tau)$. 
\\ \indent
However, what we have to calculate is the free component, which is the source of the observables. 
Eqs. (\ref{free}) give us the evolution of the $e_f$ and $n_f$, and one can easily check that these 
two equations are equivalent with the following equation for the free component:
\beq
\frac{d f^f}{d \tau}=-\frac{f^f}{\tau}+\frac{f^i}{\tau_{FO}} \left( \frac{L}{L+\tau_1-\tau} \right)\,.
\eeq{ffree}
The measured post FO spectra are given by $f^f(L+\tau_1)$.

\section{Results from the model}

Aiming for a qualitative illustration of the FO process we show below the results for the ideal massive pion gas with J\"uttner equilibrated distribution \cite{Juttner}: 
\beq
f^i(\tau,\vec{p})=\frac{g}{(2\pi)^3}e^{-\sqrt{|\vec{p}|^2+m_\pi^2}/T_i(\tau)}\,,
\eeq{Jut}
where the degeneracy of pion is $g=3$, while the baryon chemical potential in case of pions is zero. 
\\ \indent
Contrary to the illustrative example in \cite{Bjorken_FO} here we do not neglect the pion mass. 
During FO the temperature of the interacting component decreases to zero, so at late stages 
of the FO process this new calculation is better justified. 
We will see below that $T_i$ falls below $m_\pi$ quite soon, and so the J\"uttner distribution is a good approximation of the proper Bose pion distribution. 
\\ \indent
For our system we have the following EoS:
\beq
e^i=\frac{3g}{2\pi^2}m^2T_i^2K_2(a)+\frac{g}{2\pi^2}m^3T_i K_1(a)\,, \quad P^i=\frac{g}{2\pi^2}m^2T_i^2K_2(a)\,,
\eeq{EoS}
where $K_n$ is Bessel function of the second kind, and $a=m/T_i$.

The first eq. of system (\ref{int}), gives the following equation for the evolution of the temperature of the interacting component:
$$
\frac{dT_i}{d \tau}= -\frac{T_i}{\tau} \frac{4 T_i^2 K_2(a) + m T_i K_1(a) }
{12 T_i^2 K_2(a) + 5 m T_i K_1(a) + m^2 K_0(a)}
$$
\beq
- \frac{T_i}{\tau_{FO}} \left(\frac{L}{L+\tau_1-\tau} \right)\frac{3 T_i^2 K_2(a) + m T_i K_1(a) }
{12 T_i^2 K_2(a) + 5 m T_i K_1(a) + m^2 K_0(a)}\,.
\eeq{T_eq}
Furthermore, we have used the following values of the parameters: $\eta_R=4.38$, $A_{xy}=\pi R_{Au}^2$, where $ R_{Au}=7.685$ fm is the $Au$ radius, $\tau_0=0.05\ fm$, $T_i(\tau_0)=820\ MeV$, $\tau_1=5\ fm$, what leads to $T_i(\tau_1)=T_{FO}=180\ MeV$, and  $\tau_{FO}=0.5\ fm$. During the pure Bjorken case the evolution of the temperature is govern by a simplified eq. (\ref{T_eq}), without the second freeze out term on r.h.s. 

In Fig. \ref{fig1} we present the evolution of the temperature of the interacting matter, $T_i(\tau)$, for different values of FO time $L$. 

\begin{figure}[htb!]
\centering
\includegraphics[width=7.5cm, height =6.0cm]{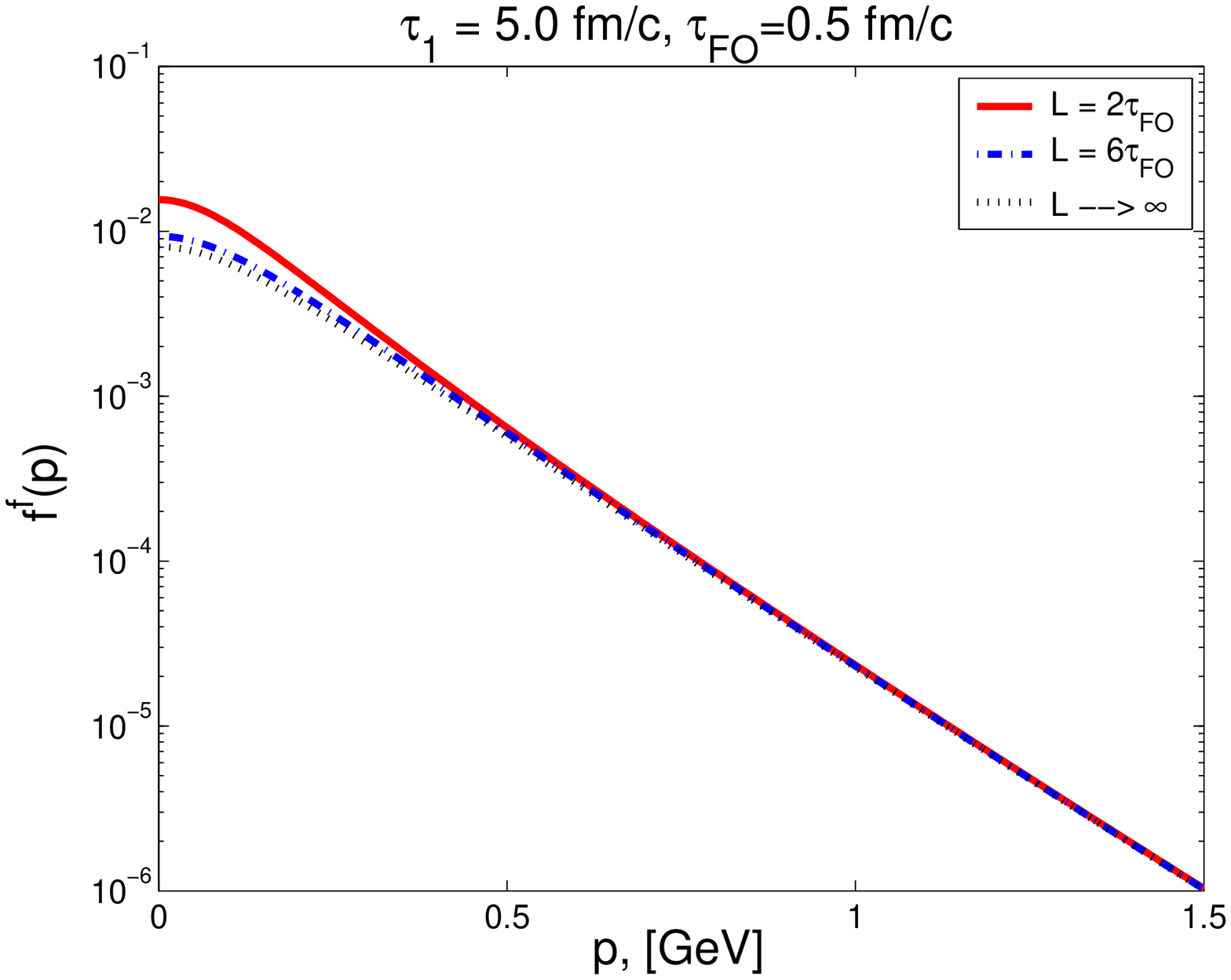} 
\includegraphics[width=5.0cm, height =5.55cm]{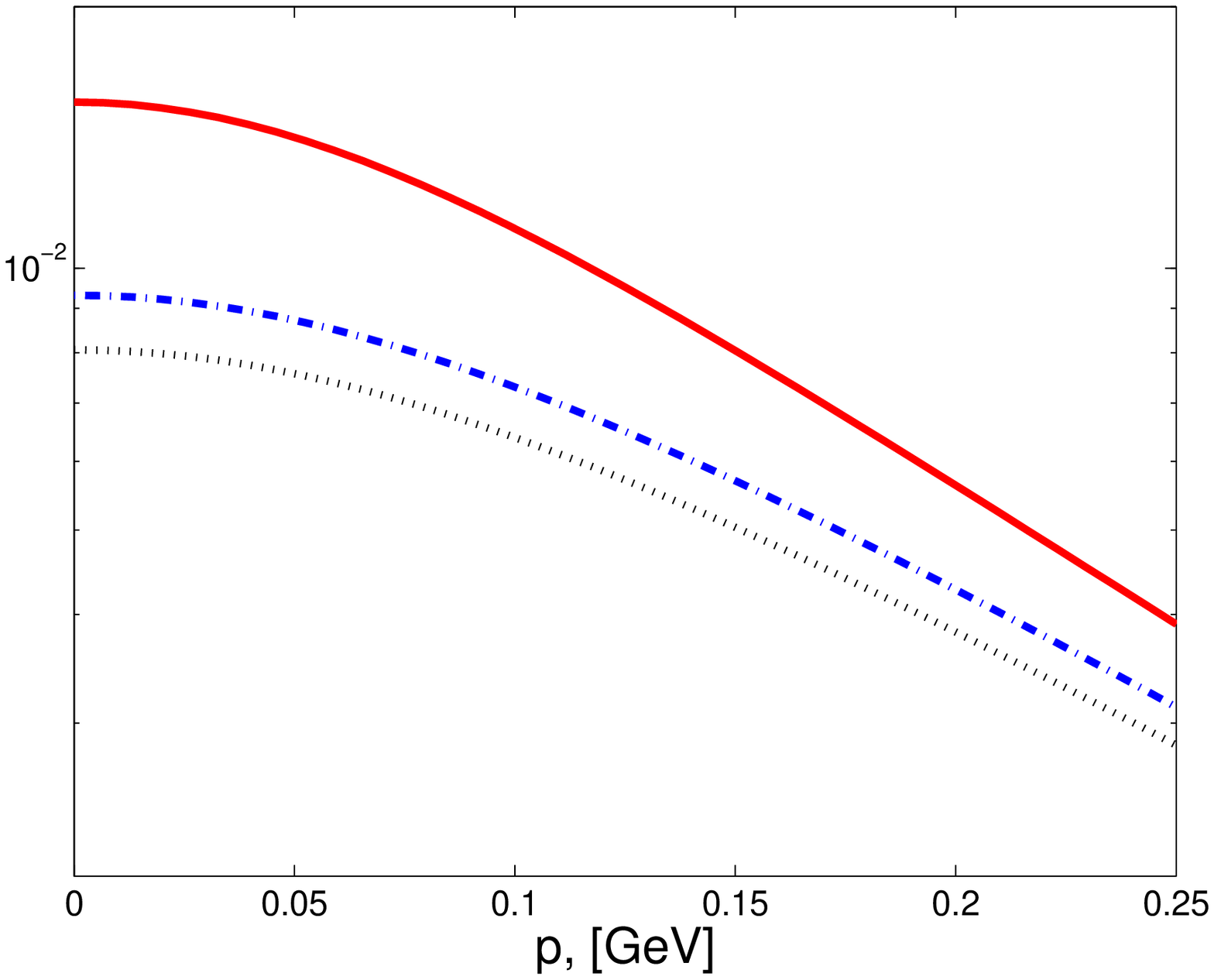}
\caption[]{Final post FO distribution for different FO layers as a function of the momentum in the
FO direction, $p=p^x$ in our case ($p^y=p^z=0$). The initial conditions are specified in the text.}
\label{fig2}
\end{figure}

\begin{figure}[htb!]
\centering
\includegraphics[width=8.0cm, height =6.15cm]{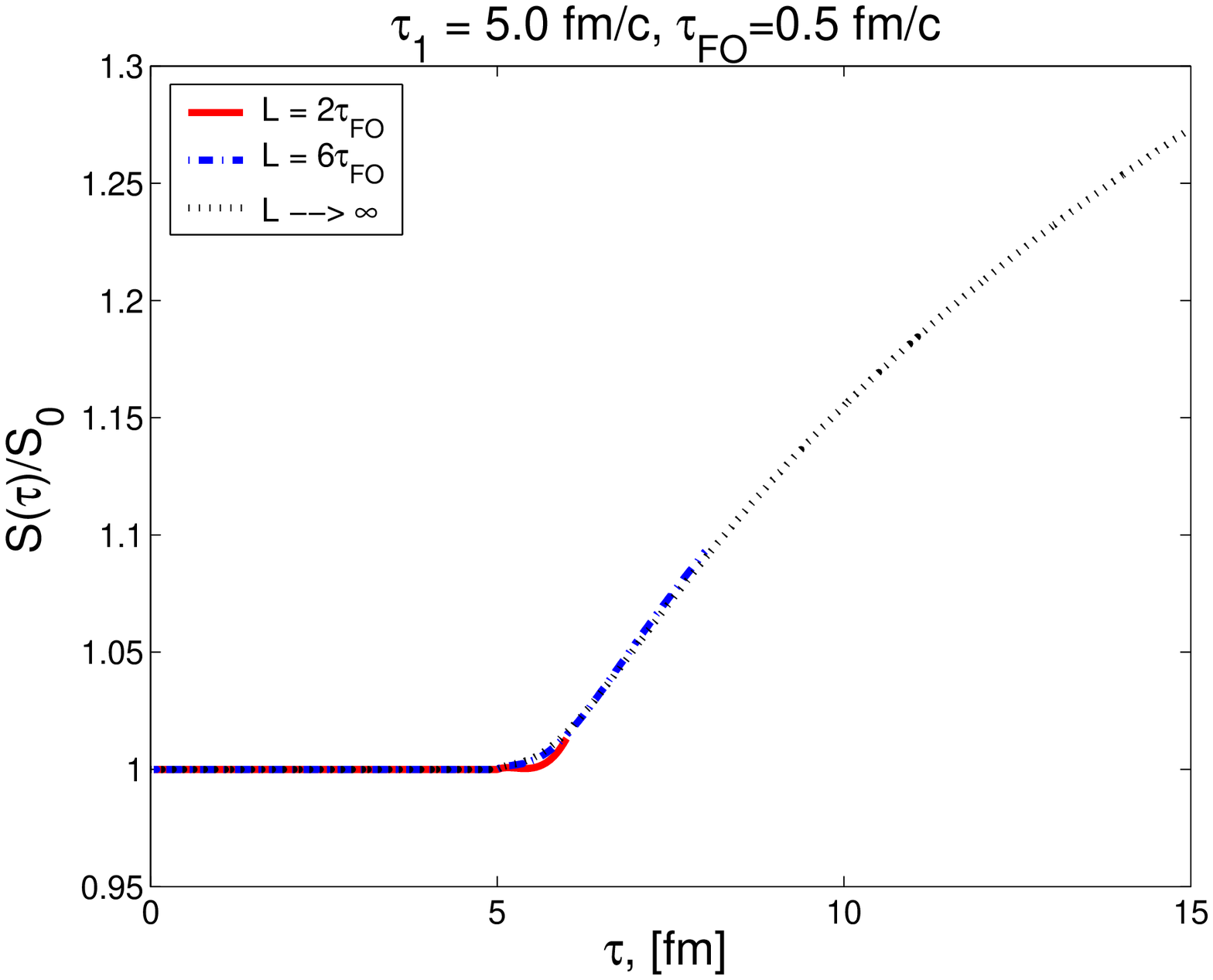} 
\includegraphics[width=4.5cm, height =5.7cm]{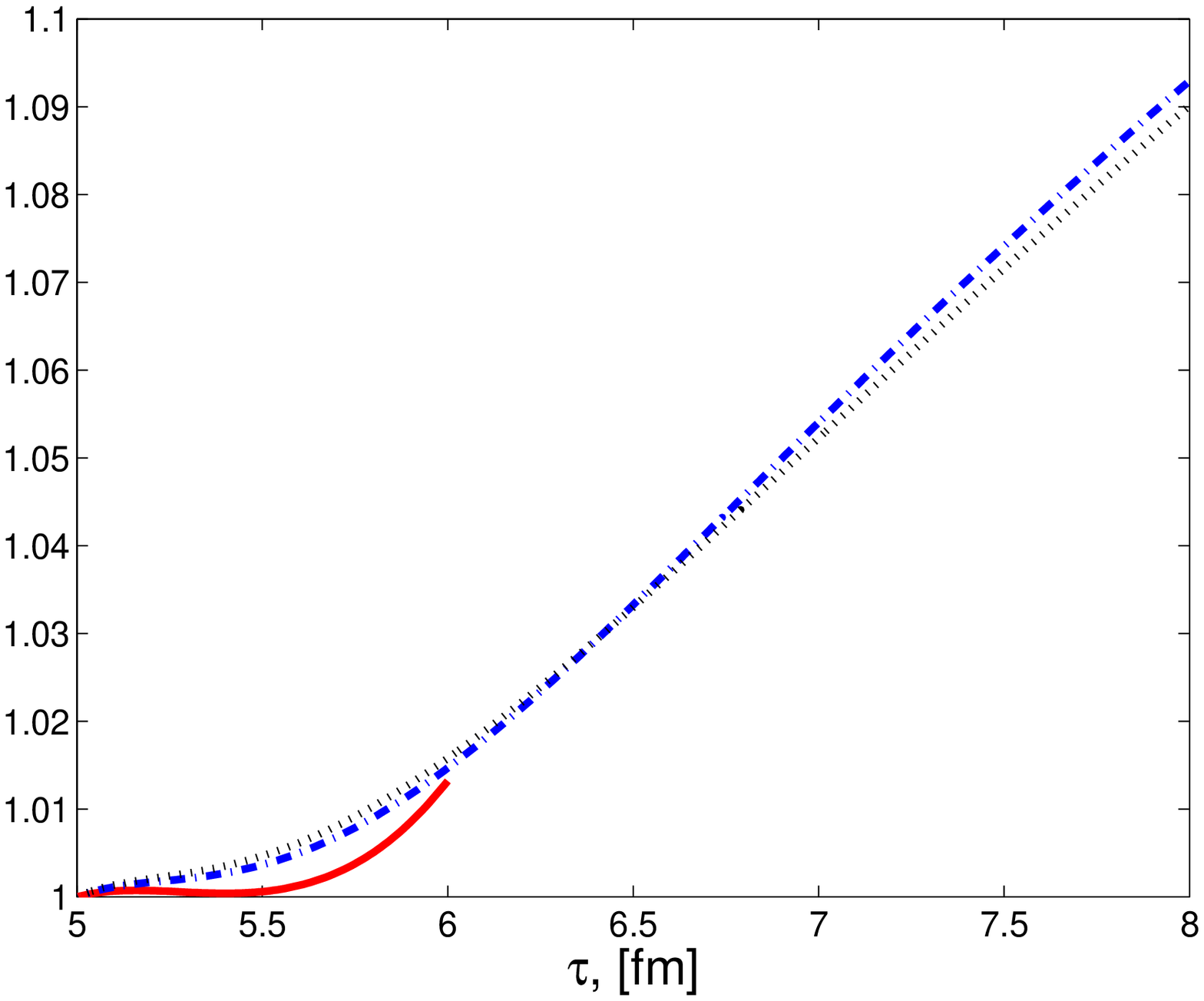}
\caption{Evolution of the total entropy for
different FO layers.  The initial conditions are specified in the text. }
\label{entrop}
\end{figure}

As it was already shown in Ref.\cite{old_TL_FO,Mo05b}, the final post FO particle distributions, shown on Fig. \ref{fig2},
are non-equilibrated distributions, which deviate from thermal ones particularly in the low momentum region.
By introducing  and varying the thickness of the FO layer, $L$, we are strongly affecting the evolution
of the interacting component, see Fig. \ref{fig1}, but the final post FO distribution shows strong universality:
for the FO layers with a thickness of several $\tau_{FO}$ post FO distribution already looks very close to that for an infinitely long FO calculations, see Fig. \ref{fig2} left plot. Differences can be observed only for the very small momenta, as shown in  Fig. \ref{fig2} right plot.
So, the inclusion of the expansion into our consideration does not smear out this very important feature of the 
gradual FO.
\\ \indent
It is important to always check the non-decreasing entropy condition \cite{Bjorken_FO,cikk_2} to see whether such 
a process is physically possible.
Figs. \ref{entrop} present the evolution of the total entropy, $S(\tau)$, calculated based on the full 
distribution function, $f(\vec{p})=f^i(\vec{p})+f^f(\vec{p})$:
\beq
s(\tau)=\int d^3 p f(\tau) \left[ 1 - \ln \left( \frac{(2\pi)^3}{g} f(\tau) \right)\right]\,, \quad S(\tau)=s(\tau)V(\tau)\,.
\eeq{stot}
During pure Bjorken phase total entropy remains constant, as expected, but during phase II it constantly 
increases until FO is finished.

\indent
From this figure we can make an important conclusion, that gradual freeze out produces entropy.
For the late stages of the FO this can be approximated analytically.
\beq
\frac{ds(\tau)}{d \tau}=\int d^3 p \frac{df(\tau)}{d\tau} \left[ 1 - \ln \left( \frac{(2\pi)^3}{g} f(\tau) \right)\right] - \int d^3 p  \frac{df(\tau)}{d\tau} \,, 
\eeq{ds}
At the late stages of the FO contribution of the interacting component can be neglected, so $f\approx f^f$, and, from eq. (\ref{ffree}), $\frac{d f^f}{d \tau}\approx-\frac{f^f}{\tau}$. So for the late stages of the reaction we have the following equation for the entropy evolution:
\beq
\frac{ds(\tau)}{d \tau}\approx -\frac{s(\tau)}{\tau}  + \frac{n^f(\tau)}{\tau}\,.
\eeq{ds2}
And since in the Bjorken geometry $V(\tau)\sim \tau$ we obtain:
\beq
\frac{dS}{d \tau}=\frac{d\left(s V\right)}{d \tau } \approx -\frac{s V}{\tau}  + \frac{n^f V}{\tau} + \frac{s V}{\tau} = \frac{N_f}{\tau}\,,
\eeq{ds_fin}
where $N_f$ is total number of frozen out particles. 
Thus, we see that total entropy increases during the simultaneous expansion and gradual FO, and at the very
end of the FO process it increases logarithmically.

\section{Conclusions}

In this paper we presented a gradual FO model including a Bjorken like expansion, being an 
extension to the older  versions \cite{old_SL_FO,old_TL_FO,Mo05a,Mo05b}, which allowed us 
to study FO in a layer of any thickness, $L$, from $0$ to $\infty$.
Another important feature of the proposed model is that it connects the pre FO hydrodynamical quantities, 
like energy density, $e$, baryon density, $n$, with post FO distribution function in a relatively simple way, 
and furthermore allows analytical analysis for simple systems, like massless pion gas \cite{Bjorken_FO}.
\\ \indent
The results show that the inclusion of the expansion into FO model, although strongly affects the evolution
of the interacting component, does not smear out the universality of the final post FO distribution, observed already in Refs. \cite{old_TL_FO,Mo05a,Mo05b}.
\\ \indent
Another important conclusion of this work, stressing once again the importance to always check the 
non-decreasing entropy condition  \cite{Bjorken_FO,cikk_2}, is that long gradual freeze 
out may produce substantial amount of entropy, as shown on Fig. \ref{entrop}.

\end{document}